\documentclass[a4paper,fleqn,usenatbib]{mnras}

\usepackage{newtxtext,newtxmath}

\usepackage[T1]{fontenc}
\usepackage{ae,aecompl}
\usepackage{graphicx}	
\usepackage{amsmath}	
\usepackage{amssymb}	
\usepackage{hyperref}
\usepackage{microtype}
\usepackage{verbatim}
\usepackage{graphicx}
\usepackage{url}
\usepackage{caption}
\usepackage{multicol}
\usepackage{soul}

\title[Discovery of pulsations in the ULX NGC 1313 X-2]{\centering{The discovery of weak coherent pulsations in the ultraluminous X-ray source NGC 1313 X-2}}
\author[R. Sathyaprakash et al.]{R. Sathyaprakash,$^{1}$\thanks{E-mail: Rajath.Sathyaprakash@durham.ac.uk}
T.\,P. Roberts$^{1}$,
D.\,J. Walton$^{2}$,
F. Fuerst$^{3}$,
M. Bachetti$^{2}$,
C. Pinto$^{4}$,
\newauthor
W.\,N. Alston$^{2}$,
H.\,P. Earnshaw$^{5}$,
A.\,C. Fabian$^{2}$,
M.\,J. Middleton$^{6}$,
R. Soria$^{7}$$^{,}$$^{8}$
\\
$^{1}$Centre for Extragalactic Astronomy, Durham University, Department of Physics, South Road, Durham DH1 3LE, UK\\
$^{2}$Institute of Astronomy, University of Cambridge, Madingley Road, Cambridge CB3 0HA, UK\\
$^{3}$European Space Astronomy Centre (ESAC), Science Operations Department, 28692, Villanueva de la Ca\~{n}ada, Madrid, Spain\\
$^{4}$ESTEC/ESA, Keplerlaan 1, 2201AZ Noordwijk, The Netherlands\\
$^{5}$Cahill Center for Astronomy and Astrophysics, California Institute of Technology, Pasadena, CA 91125, USA\\$^{6}$Department of Physics and Astronomy, University of Southampton, Highfield, Southampton SO17 1BJ, UK\\$^{7}$College of Astronomy and Space Sciences, University of the Chinese Academy of Sciences, Beijing 100049, China\\$^{8}$Sydney Institute for Astronomy, School of Physics A28, The University of Sydney, Sydney, NSW 2006, Australia}

\date{Accepted XXX. Received YYY; in original form ZZZ}

\pubyear{2019}

\begin{document}
\label{firstpage}
\pagerange{\pageref{firstpage}--\pageref{lastpage}}
\maketitle
\begin{abstract}
\begin{centering}
We report the detection of weak pulsations from the archetypal ultraluminous X-ray source (ULX) NGC 1313 X-2. Acceleration searches reveal sinusoidal pulsations in segments of two out of six new deep observations of this object, with a period of $\sim 1.5$~s and a pulsed fraction of $\sim 5\%$.  We use Monte Carlo simulations to demonstrate that the individual detections are unlikely to originate in false Poisson noise detections given their very close frequencies; their strong similarity to other pulsations detected from ULXs also argues they are real.  The presence of a large bubble nebula surrounding NGC 1313 X-2 implies an age of order 1~Myr for the accreting phase of the ULX, which implies that the neutron star's magnetic field has not been suppressed over time by accreted material, nor has the neutron star collapsed into a black hole, despite an average energy output into the nebula two orders of magnitude above Eddington.  This argues that most of the accreted material has been expelled over the lifetime of the ULX, favouring physical models including strong winds and/or jets for neutron star ULXs.
\end{centering}
\end{abstract}


\section{Introduction}

\begin{figure*}
  \begin{minipage}{\textwidth}
    \centering
    \includegraphics[width=0.85\textwidth]{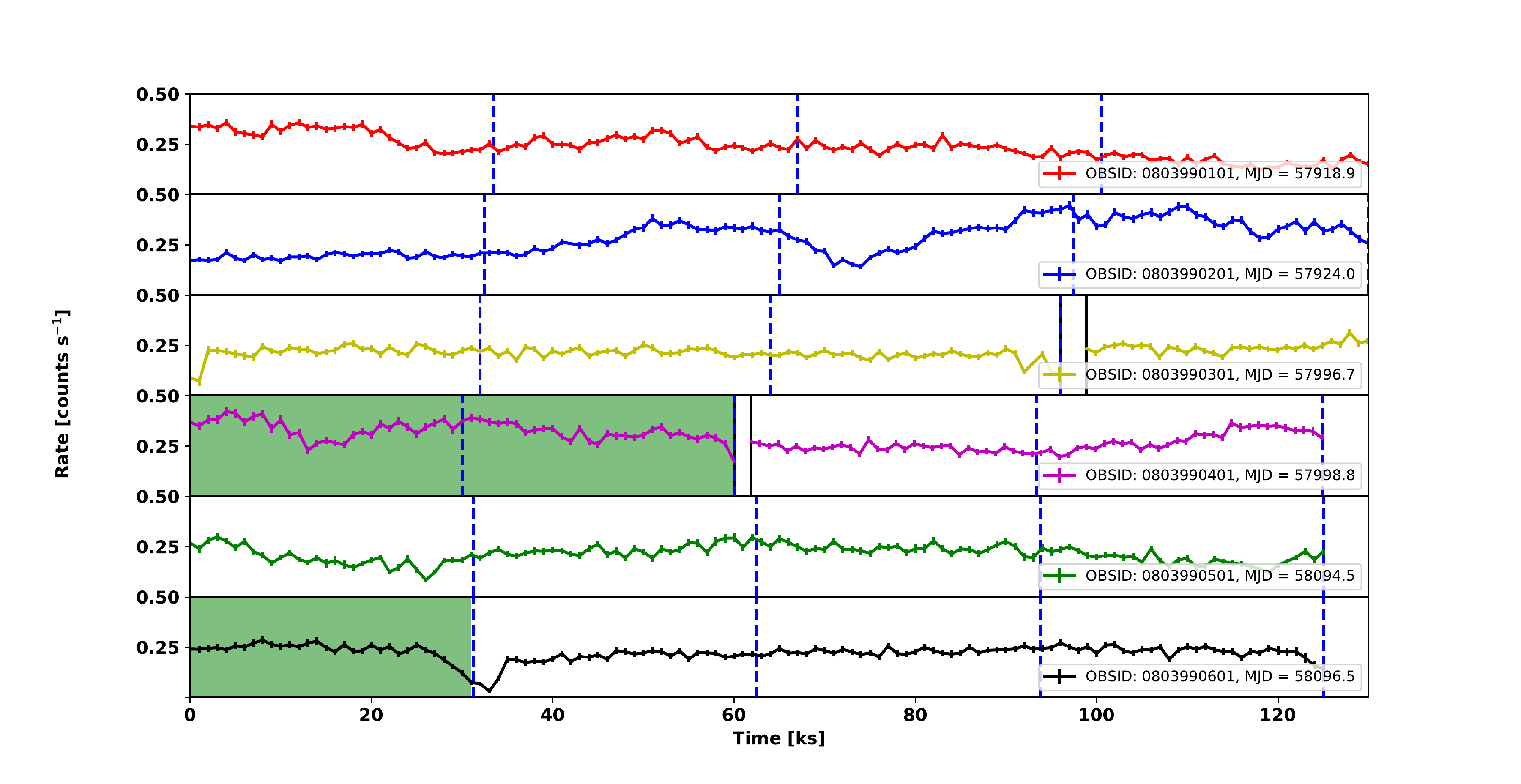}
    \caption{The EPIC-pn light curves of the six new, deep \textit{XMM-Newton} observations of NGC 1313 X-$2$ in time bins of 1000 seconds. In the analysis each light curve is split first into (i) four segments of $\sim$ 30 ks exposure length (indicated by the dashed blue lines) and then into (ii) two longer segments of $\sim$ 60 ks exposure length, with the acceleration search run on each individual segment. The shaded green regions correspond to epochs during which the pulsations were detected, and the black lines separate the non-contiguous segments (owing to a pause in exposure) of the third and fourth observations. The legend specifies the observation identifier and the time of the mid-point of each exposure (in Modified Julian Date).}
    \label{Fig:lightcurves}
  \end{minipage}\\[1em]
\end{figure*} 

Ultraluminous X-ray sources (ULXs) are extragalactic off-nuclear X-ray sources with apparent X-ray luminosities that match or exceed the Eddington limit for a 10 $M_{\odot}$ black hole (see \citealt{1} for a recent review).  For many years the debate over ULXs focussed on the mass of the black hole powering this phenomenon (e.g. \citealt{2}; \citealt{3}; \citealt{4}).  However the detection of coherent pulsations in four ULXs implies that each must host a neutron star (NS) primary radiating at up to several hundred times its Eddington limit (\citealt{5}; \citealt{6a, 6b}; \citealt{8}; \citealt{9}).  Hence, the key questions for ULXs now revolve around what fraction of the population host NSs, and the physics of super-Eddington accretion in strong magnetic fields.  The discovery of further NS-ULXs is crucial to answering both questions, with signatures such as cyclotron resonant scattering features and possible transitions in and out of a propeller regime revealing more NS-ULX candidates (e.g. \citealt{10}; \citealt{11}).  Indeed, broad-band spectroscopic similarities in a small sample of ULXs suggest NSs may be ubiquitous in the ULX population (\citealt{12}).
This can only be confirmed by more evidence for the presence of NSs in individual objects.  This evidence is growing with, for example, the first detection of a Galactic pulsating ULX (PULX, e.g. \citealt{13}).  Here we report a new detection of pulsations from a hitherto archetypal extragalactic ULX, NGC 1313 X-2.  This PULX is located on the outskirts of the barred spiral galaxy NGC 1313, and is historically amongst the best studied ULXs in terms of both its X-ray properties (e.g \citeauthor*{14} \citeyear{14}; \citeauthor*{16} \citeyear{16}) and its optical counterpart and environment (e.g. \citeauthor*{17} \citeyear{17}; \citealt{19}).  In particular, a lack of radial velocity variations in its optical spectrum (\citealt{20}) and the patterns of change in its hard X-ray spectrum and flux variability could suggest it is viewed at a low inclination angle (\citealt{21}; \citealt{22}), affording a direct view of the central regions of its accretion flow.

\section{Observations and data analysis} 
The barred spiral galaxy NGC 1313 was observed by \textit{XMM-Newton} seven times between 2017 June 16 and 2017 September 24, as part of a Large Programme observing both ULXs in the galaxy: NGC 1313 X-1 and X-2 (see Pinto et al. in prep. for a summary of the data obtained). The latter is the target of this study, and it lies $\sim$ 7 arcminutes off-axis from the nominal aim-point of the observations. The observations were carried out with the EPIC-pn and MOS detectors in full-frame imaging mode. The raw observation data files (ODF) were processed with the Science Analysis Software ({\sc sas})v.16.  We excluded observation 0803990701 from further analysis at this stage as it contained only $\sim 15$~ks of exposure; all other observations were exposed for $> 120$~ks.  Source events were extracted in a circular aperture of 25 arcsecond radius centred on the target, while the background region was chosen to be adjacent to the source on the same chip. We then used the the {\sc sas} task {\tt{evselect}} to derive the background-subtracted 0.3-10 keV light curves in 73~ms time bins, which corresponds to the maximal time resolution of the EPIC-pn camera (in full-frame mode). We did not utilise data from the MOS detectors, since they do not have sufficient time resolution to be able to detect pulsations with frequencies larger than $0.2$~Hz. We implemented the standard filtering of events as recommended in the \textit{XMM-Newton} science threads. Namely, we selected only single or double patterned events ({\tt{PATTERN}} $<= 4$) and used the standard flagging criteria (i.e. {\tt{\#XMMEA\_EP}}) that ensures the exclusion of hot pixels and events falling outside the field of view. In some observations the source emission was spread over more than one chip, and in such circumstances, we combined the data from both chips to construct the light curves.


\subsection{Discovery of coherent pulsations}

\begin{figure*}
  \begin{minipage}{\textwidth}
    \centering
    \includegraphics[width=0.85\textwidth]{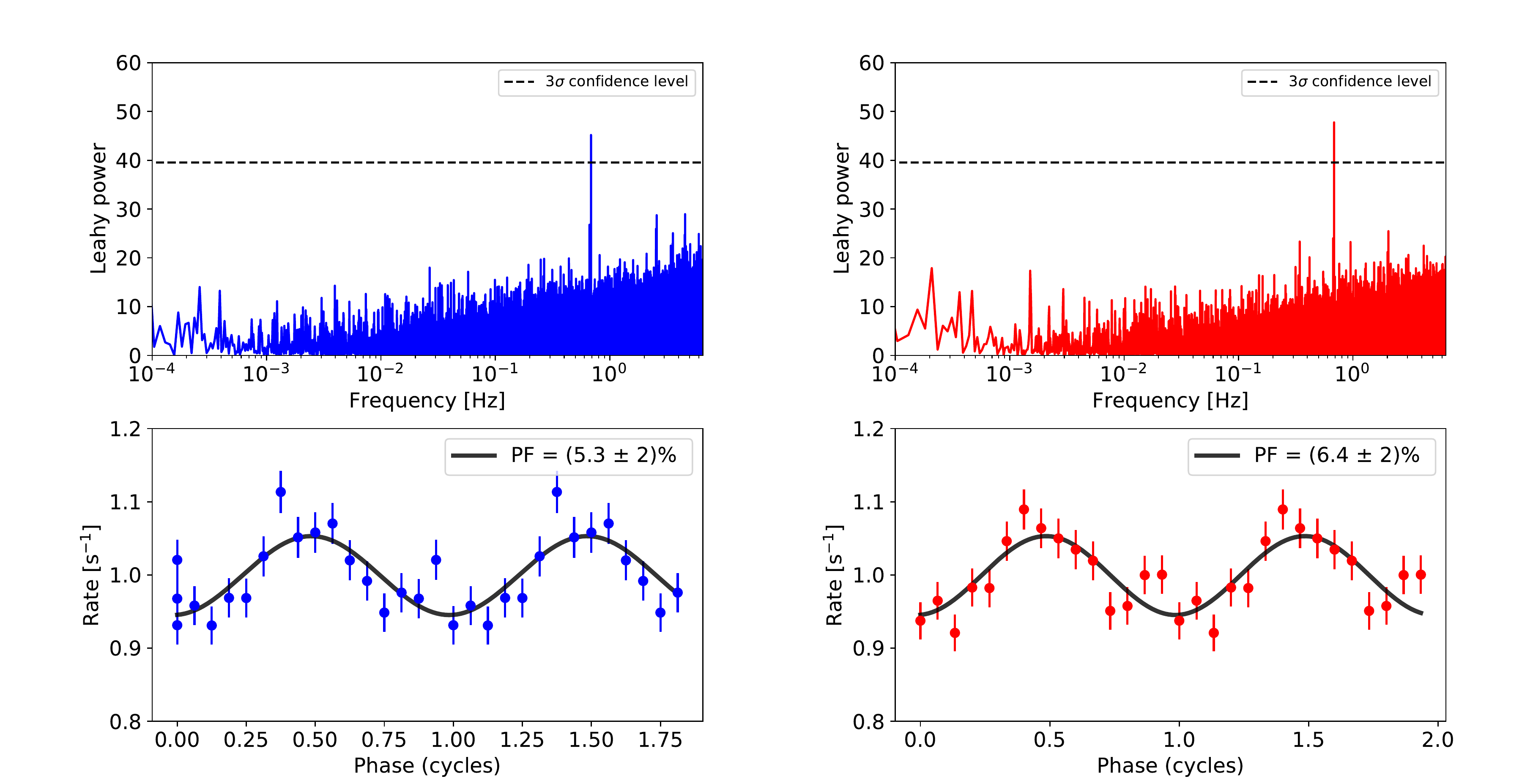}
    \caption{(\textit{Top row}) The power spectral densities of the two \textit{XMM-Newton} observations during which coherent pulsations were detected, after correction for the measured accelerations. The dotted line overlaid on each plot shows the 3$\sigma$ confidence level, after taking into account the number of trials. (\textit{Bottom row}) The pulse profiles of the corresponding observations folded on the period and period derivative quoted in Table~\ref{Tab:pulses}.}
    \label{Fig:PSDs}
  \end{minipage}\\[1em]
\end{figure*} 

As the starting point of our timing analysis, we converted the event times from the EPIC-pn detector to the barycentre of the Solar System with the {\sc sas} task {\tt{barycen}}, using the \textit{Chandra} source position (RA = 49.594 degrees, DEC = -66.601 degrees) and the DE-200 Solar System ephemeris (\citealt{23}).  We computed the power spectra of the six remaining observations using the {\sc heasarc} (v6.22) tool {\tt{powspec}}. Initially, these did not indicate any signals peaking above the $1\sigma$ white-noise threshold (\citealt{24}). However, if the source has a relatively strong period derivative $\dot{f}$, this causes the signal to drift across multiple Fourier bins in the power spectrum, with the number of bins scaling as $\dot{f}T_{\text{obs}}^2$ for observations of length $T_{\text{obs}}$, thus reducing the amplitude of the signal power and the sensitivity of any given search. In order to correct for such an effect, we implemented the Fourier domain acceleration search method by using the {\sc presto} software (\citeauthor*{25} \citeyear{25}). This method is optimised if the acceleration (or the period derivative) of the source is constant throughout the observation, which is best satisfied if the exposure time is smaller than a tenth of the orbital period of the system. Therefore, we split the light curve of each observation, first into (i) contiguous 30~ks segments, which accounts for the strongest period derivative measured so far in PULXs (i.e. $\sim 10^{-7}$ Hz s$^{-1}$; \citealt{9}), and later into (ii) larger 60~ks segments, which allows for a stronger detection of candidates with smaller period derivatives. We show the full light curves, and their segmentation, in Figure~\ref{Fig:lightcurves}; in all cases we ran the search separately on each segment. 

\begin{figure*}
  \begin{minipage}{\textwidth}
   \includegraphics[scale=0.55]{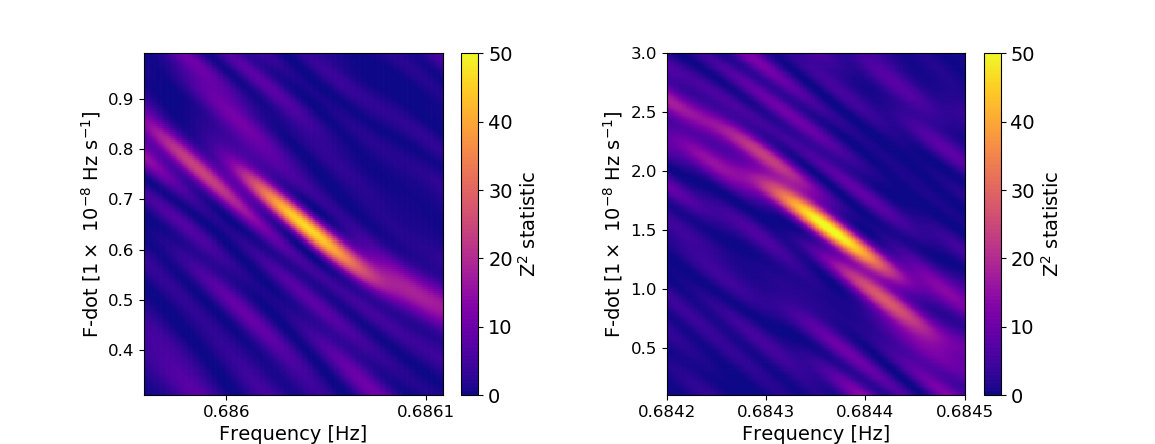}  
    \centering	  
    \caption{Epoch-folding search results for the two observations revealing pulsations. This re-affirms the results obtained through {\sc presto}. The colour bar shows the Z$^{2}$ statistic with peak values being above the 3$\sigma$ significance level (i.e. $> 40$).}
    \label{Fig:presto}
\end{minipage}\\[1em]
\end{figure*}

This resulted in the possible detection of coherent pulsations in a 30~ks and a 60~ks segment of two separate observations, out of the total of six, indicating that NGC 1313 X-2 is a new addition to the class of PULXs. Both of these detections were found to occur at similar Fourier frequencies and with a statistical significance larger than 3$\sigma$ (see Figures~\ref{Fig:PSDs} and \ref{Fig:presto})\footnote{Here, we refer to the significance reported by the {\tt{accelsearch}} routine in {\sc presto}, which accounts for the total number of trials}. We folded the light curves of these two segments on the best-fit values of the pulse frequency and its derivative (determined using {\sc presto}) in order to obtain the pulse profiles (\citealt{26}; Figure~\ref{Fig:PSDs}). These appear sinusoidal, as per other PULXs, but the pulsed fractions at $\sim 5\%$ are notably smaller than previously seen (e.g. \citealt{5}; \citealt{8}). We further verified our timing analysis through an epoch folding search on the two light curve segments shown in Figure~\ref{Fig:lightcurves} by using the {\sc hendrics} tool {\tt{HENzsearch}} (\citealt{27}). Here, a test for periodicity was performed by evaluating the Z$^{2}$ statistic over a grid of $f - \dot{f}$ values (c.f. \citeauthor{39} \citeyear{39} for further details). The results of the test are shown in Figure~\ref{Fig:presto}, with the peak Z$^{2}$ values in both observations being larger than the $3\sigma$ significance level, consistent with the results obtained through {\sc presto}. 

We calculated the unabsorbed 0.3-10 keV X-ray luminosity for each observation by fitting a simple power-law plus multi-colour disc model to the EPIC-pn spectra in {\sc xspec} and correcting for both the Galactic and intrinsic absorption columns. The X-ray spectra were extracted by following the standard procedure outlined in \citet{28}. We tabulate these luminosities alongside the characteristics of the detected pulses in Table~\ref{Tab:pulses}. Finally, we note that we do not detect pulsations in any archival observations with similar quality of data and exposure as in the six observations from our programme.  

\begin{table*}
\caption{Characteristics of the detected pulsations.}
\centering
\begin{tabular}{c c c c c c c}
\hline
Data set & Date$^{\text{a}}$ & $f^{\text{b}}$ & $\dot{f^{\text{c}}}$ & {\tt{PRESTO}} significance & Pulsed fraction & $L_{\text{X}}^{\text{d}}$ \\[0.5ex]
\newline 
\newline
Obs.ID & (MJD) & (Hz) & (Hz s$^{-1}$) & ($\sigma$) & (\%) & (10$^{40}$ erg s$^{-1}$) \\
\hline
0803990401 & 57998.0 & 0.686041(1) & 6.5(5) $\times 10^{-9}$ & 3.00 & 5.3 $\pm$ 2 & 1.99 \\
0803990601 & 58096.0 & 0.6843642(6) & 1.5(2) $\times 10^{-8}$ & 4.12 & 6.4 $\pm$ 2 & 1.44 \\
\hline
\end{tabular}
\begin{minipage}{0.8\textwidth}
Notes:  $^{\text{a}}$ mid-point of the observation segment;
	$^{\text{b}}$ detected frequency;
	$^{\text{c}}$ detected frequency derivative;
	$^{\text{d}}$ intrinsic X-ray luminosity in units of 10$^{40}$ erg s$^{-1}$ in the 0.3-10 keV band.
\end{minipage}
\label{Tab:pulses}
\end{table*}

\subsection{Verifying the detection significance with Monte Carlo simulations}

In \citet{24}, the significance level for the detection of a coherent signal (above a given threshold) is evaluated by considering the distribution of the noise powers and the number of Fourier bins in the power spectrum. The significance given by {\sc presto} is calculated in the same manner. However, it does not consider the fact that we split each light curve into several segments, and also searched for signals dispersed over an initially unknown number of Fourier bins $z_{\text{max}}$. Searching over a range of different $z_{\text{max}}$ values for each segmented light curve vastly increases the parameter space, making it more likely to obtain a false positive. Therefore, Monte Carlo simulations are required in order to account for these factors and derive an estimate of the true significance of our detections. 

We simulated 10000 fake light curves containing pure Poisson noise with exposure times set to match those of the real observations, and mean count rates sampled randomly from the range of count rates observed in archival {\it XMM-Newton\/} and {\it Swift\/} observations of NGC 1313 X-2 (with the latter converted to equivalent count rates in {\it XMM-Newton\/} using {\sc pimms}).  Each light curve was split in the same manner as the real data (including the addition of GTI timing gaps), and was searched for pulsations using {\sc presto} as described in Section 2.1.  We then picked 12 simulated light curves at random to match the 12 long observations in the actual data (including archival observations). We looked for two detections of pulsations within these simulated observations with statistical significances equal to or larger than those measured for the real data, and detected pulse frequencies differing by no more than the difference between our two detections (i.e. $< 1.5 \times 10^{-3}$ Hz).  We repeated this random sampling of our simulated light curves 10000 times.  We found a false positive rate of 1 in 10000, implying that our detections are significant to more than $3.5 \sigma$ even after accounting for the "look elsewhere" effect.           


\subsection{Constraints on orbital parameters}

\begin{figure}
   \centering
    \includegraphics[width=9.5cm]{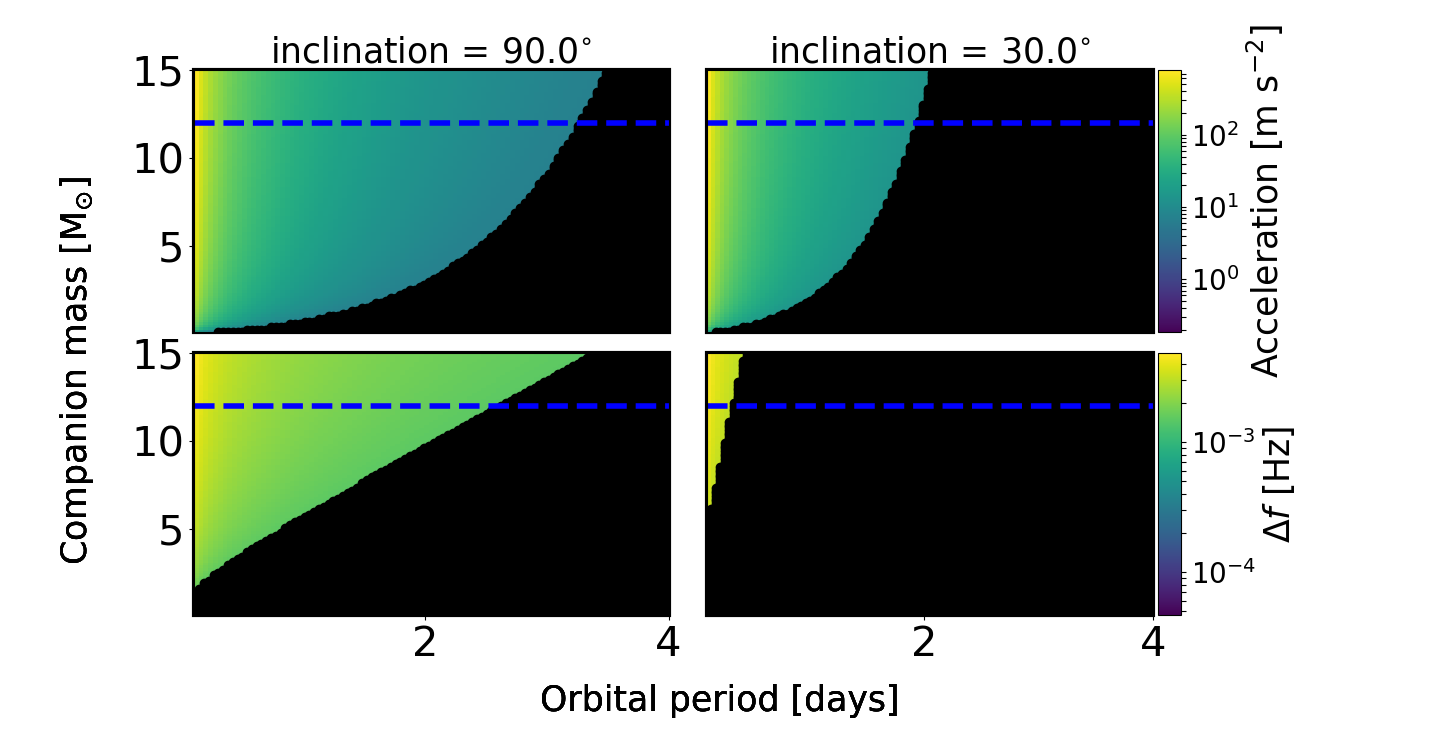}
    \vspace{-12pt}
    \caption{Constraints on the orbital parameters of NGC 1313 X-2.  In the top panel, we plot the expected line-of-sight acceleration (shown by the colour-bar) of a 1.4 $M_{\odot}$ neutron star in a circular orbit around a companion with varying masses and orbital periods. In the bottom panel, the colour-bar indicates the expected pulse frequency shift in a single orbital cycle (in Hz) as a function of the companion mass and orbital period. In both cases, the black shaded regions correspond to an area of parameter space that does not give rise to accelerations (top-panel) or pulse frequency shifts (bottom-panel) as large as the value actually observed (see Table~\ref{Tab:pulses}), and are therefore ruled-out. The blue dashed line marks the upper limit on the companion star mass inferred through stellar population studies by \citet{29}.}
    \label{Fig:orbits}
\end{figure}

In order to set preliminary constraints on the orbital parameters of the system, we will conservatively assume that any variation in the pulsation frequency is dominated by an orbital modulation (although see Section 3). In this case, the line-of-sight acceleration of the source $\alpha$ can be expressed in terms of the spin parameters as $\alpha = c(\dot{f}/f$).  The value of $\dot{f}$ measured in the sixth observation is $1.6(2) \times 10^{-8}$~Hz~s$^{-1}$, which corresponds to $\alpha = 7(1)$ m s$^{-2}$.  In order to produce such a large acceleration, the companion mass $M_{c}$ and the orbital period $P_{\text{orb}}$ must be restricted to the range of values illustrated in the top panel of Figure~\ref{Fig:orbits}. Moreover, if the observed pulse frequency shift between the two observations is solely due to the orbital motion of the source, this imposes further limits on the orbital parameters (as illustrated in the bottom panel of Figure~\ref{Fig:orbits}). In deriving these estimates we have assumed that the pulsations indicate the presence of a NS primary in the binary system with a mass $\sim 1.4 M_{\odot}$. We also assume a circular orbit, which appears to be the case for several other PULXs likely due to the action of strong tidal torques (\citealt{5}, \citealt{6a}, \citealt{8}). We consider the constraints on the orbital parameters for two separate inclination angles, and find that it is particularly difficult to explain the observed frequency shift if the system is viewed moderately face-on (i.e. $<$ 30 degrees; second figure in the bottom panel). However, since the source is classified to be hard ultraluminous from its X-ray spectrum, its central regions are likely to be seen close to face-on rather than edge-on (Sutton, Roberts \& Middleton 2013). This apparent contradiction can be reconciled if the spin and orbital angular momenta of the system are misaligned. 
Further, \citet{29} have derived an upper limit on the mass of the companion star from studies of the young OB association in its vicinity. By fitting stellar isochrones to the {\it HST\/} optical magnitudes of these stars, they were able to measure the cluster age to be $\sim$ 20 Myrs. This limits the maximum mass of the companion star of the ULX to be $\lesssim 12 M_{\odot}$, provided that it is associated with the star cluster. A typical B-type companion consistent with this mass limit has a radius of $\sim 4 R_{\odot}$ and must overflow its Roche-Lobe during the ULX phase. In such a scenario, the period-density relation (Frank et al. 2002) independently constrains the orbital period, which we find to be consistent with the range of values inferred from X-ray timing (provided that there is some spin-orbit misalignment). Finally, we can use the constraints derived from this analysis to verify that our definition of a false positive in the previous section (i.e. any two detections with pulse frequencies differing by $\sim 1.5 \times 10^{-3}$ Hz or less) is not physically unreasonable. Indeed, for the allowed set of parameters shown in Figure~\ref{Fig:orbits}, the expected change in the pulse frequency in a single orbital cycle will not exceed $5 \times 10^{-3}$ Hz. If we re-run the Monte Carlo simulations by allowing the pulse frequencies to differ by as much as the latter value, we find that this does not change the false positive rate quoted in the previous section.        
\section{Discussion \& conclusions}

We report the detection of coherent pulsations in individual segments of two \textit{XMM-Newton} observations of NGC 1313 X-2 (out of a total of twelve).  The detected pulsations are notably weaker than for previous PULXs, with the measured pulsed fractions ($\sim 5\%$) being roughly a factor of two smaller.  As a result, each individual detection is only marginally significant at $\sim 3-4\sigma$; however the two detections have remarkably similar pulsation frequencies, and we confirm using Monte Carlo simulations that the chance of detecting two false positives with such similar frequencies is extremely low, so we are confident that we are observing a {\it bona fide\/} new member of the PULX class.  This is supported by the detected frequencies (that can be associated with the spin of the neutron star) and their derivatives, in addition to the sinusoidal pulse profiles, all of which are very similar to those of previously discovered sources (\citealt{5}; \citealt{6a,6b}). The pulsations are transient on intra-observational timescales, which is reminiscent of the behaviour seen in M82 X-2 (\citealt{5}), although the physical mechanism for this behaviour remains poorly understood. 

We were unable to measure an orbital modulation of the rotational frequency $f$ given the sparse number of pulsation detections. However, given the constraints on orbital acceleration provided by the frequency derivatives we were able to calculate the dependency of orbital period on mass of the companion to the NS; it is notable that we rule out periods as large as the 6.1 day period reported by \citet{30} on the basis of {\it HST\/} data for the optical counterpart to NGC 1313 X-2 (although the veracity of this period was questioned by \citealt{31} on the basis of a wider dataset).  The lack of constraints on the orbital modulation of the pulsations means that we cannot estimate secular (i.e. accretion-driven) changes in the pulsation frequency with any great certainty. Although, we can at least state that the observed frequency reduction of $\sim 1$ mHz over 100 days between the two pulse detections suggests a spin down rate $\dot{f} \sim -1.2 \times 10^{-10}$ Hz s$^{-1}$, which is notable as other PULXs all show secular spin increases. However, the instantaneous values of $\dot{f}$ measured during both epochs have the opposite sign and are significantly larger (see Table~\ref{Tab:pulses}). Therefore, we caution that the observed frequency variation may well be dominated by orbital motion.  We note that if the source is indeed spinning-down between the two observations, then this is probably not due to a transition to the propeller regime (\citealt{32}), as this is naturally accompanied by a dramatic decline in the X-ray luminosity by several orders of magnitude.  Unlike other PULXs (\citealt{33}; see also \citealt{11}) no such diminution of luminosity has been seen for NGC 1313 X-2 (cf. Figure~\ref{Fig:lightcurves}; also \citealt{34}).

NGC 1313 X-2 is unusual amongst ULXs as the presence of a large ($\sim 400$~pc diameter), shock-ionised bubble nebula around it means that we can infer an age for the active phase of the ULX. \citet*{35} calculate this age as $\sim 1$~Myr for NGC 1313 X-2.  This appears to contradict the assertion of \citet*{36} that PULXs are rare because the accretion of material onto the surface of the neutron star acts to rapidly diminish its magnetic field; NGC 1313 X-2 demonstrates that this is not the case.  Indeed, current binary evolution models show that there are only a small number of scenarios in which stable Roche Lobe overflow persists for more than $\sim 1$~Myr onto a NS (e.g. \citealt{37}).  This likely implies that NGC 1313 X-2 is close to the end of its lifetime as a ULX, and yet it is still pulsating.  As the bubble nebula requires an average input energy of $\sim 10^{40}$~erg~s$^{-1}$ over its entire lifetime to inflate, from the kinetic output of the ULX wind/jets, this also suggests that NGC 1313 X-2 has sustained extreme ($\sim 100$ times Eddington) accretion rates from its secondary stellar companion.  It is likely that the majority of this matter was expelled in radiatively driven outflows, as if this amount of matter were retained by the NS it would have accreted $\gtrsim 1 M_{\odot}$ of material and potentially collapsed to a black hole, resulting in a cessation of pulsations.  This necessity for both a strong outflow and a pulsating compact object supports the recent hybrid models for PULXs, such as described by \citet{38} and is consistent with the recent detection of outflowing material from a PULX \citep{40}.

Finally, we note that NGC 1313 X-2 is an archetype for the 'hard ultraluminous' X-ray spectral class for ULXs, in which the object is thought to be viewed directly down the wind-delimited funnel to its central engine (Sutton et al. 2013).  This observational configuration is obviously well-suited to detecting pulsations as the emission of the central regions is visible and/or beamed into an observer's line of sight; notwithstanding the transient nature of the detected pulsations, or their faintness, the advanced age of NGC 1313 X-2 as a ULX offers some encouragement that observations with sufficient depth and baseline should detect pulsations from all those ULXs with a face-on aspect, if they contain NSs.

\section*{Acknowledgements}
The authors wish to thank the referee for their useful comments that helped improve this paper. RS is grateful for the studentship from the STFC (ST/N50404X/1). TPR was funded as part of the STFC consolidated grant ST/K000861/1.  HPE acknowledges support under NASA contract NNG08FD60C.  CP is thankful for an ESA Research Fellowship, and DJW acknowledges support from an STFC ERF.  This work is based on observations obtained with {\it XMM-Newton\/}, an ESA science mission with instruments and contributions directly funded by ESA Member States and NASA.






\bsp	
\label{lastpage}
\end{document}